\documentclass[journal ]{new-aiaa}
\usepackage{placeins}
\usepackage[utf8]{inputenc}
\usepackage{textcomp}
\usepackage{graphicx}
\usepackage{enumitem}
\usepackage{setspace}
\usepackage{float}  
\graphicspath{{./}}
\usepackage{amsmath}
\usepackage[version=4]{mhchem}
\usepackage{siunitx}
\usepackage{longtable,tabularx}
\usepackage[export]{adjustbox}
\usepackage{etoolbox}
\patchcmd{\thebibliography}
  {\list}
  {\vspace{2.9pt}\list}
  {}{}
\usepackage{doi}

\hypersetup{hidelinks}
\title{{\fontsize{20}{24}\selectfont\bfseries An ArcGIS Framework for Mapping\\
Human-Centered Noise Annoyance for AAM\\
Infrastructure Planning}}

\author{{\fontsize{16}{18}\selectfont
Swapnil Saha\raisebox{0.6ex}{\fontsize{12}{18}\selectfont 1}, 
Zubin Mistry\raisebox{0.6ex}{\fontsize{12}{18}\selectfont 2}, 
and Neelakshi Majumdar\raisebox{0.6ex}{\fontsize{12}{18}\selectfont 3}}}
\affil{{\fontsize{12}{14}\selectfont\itshape
University of Arkansas, Fayetteville, Arkansas, 72701, USA}}

\begin{document}

\maketitle

\footnotetext[1]{Graduate Research Assistant, Department of Mechanical Engineering, swapnils@uark.edu}
\footnotetext[2]{Postdoctoral Fellow, Department of Mechanical Engineering, zmistry@uark.edu}
\footnotetext[3]{Assistant Professor, Department of Mechanical Engineering, neelm@uark.edu, AIAA Member}

\begin{abstract}
{\fontsize{10}{12}\selectfont
Advanced Air Mobility (AAM) represents a transformative shift in urban transportation; however, its successful implementation depends on public acceptance, with noise emerging as a primary concern for low-altitude electric vertical takeoff and landing (eVTOL) operations. Existing studies characterize eVTOL noise using acoustic metrics, such as sound pressure level (dBA) and day-night average noise level (DNL). Expressing noise impacts in terms of human response, such as the percentage of the population that is highly annoyed (\%HA), provides a more actionable basis for assessing community impact and supporting human-centered AAM planning. This study develops a Geographic Information System (GIS)-based framework to translate eVTOL acoustic outputs into \%HA-based annoyance maps for a representative medical delivery route in Northwest Arkansas. Results show that both noise and annoyance decrease with distance from the route, with the highest impacts occurring during descent, up to 58\%, followed by climb and cruise. Census population data are integrated to estimate the number of highly annoyed individuals and identify impact hotspots across the study area. These results are further combined with distance and airspace factors to evaluate alternative routes and identify balanced routing strategies. The proposed framework bridges acoustic modeling and human perception, enabling planners and policymakers to define noise-sensitive zones, evaluate route alternatives, and support socially sustainable AAM operations. 
}
\end{abstract}

\section{  Introduction}
\vspace{0.1cm}
\lettrine {E}{lectric} Vertical Take-Off and Landing (eVTOL) is an emerging domain in Advanced Air Mobility (AAM), with potential not only for transporting goods and medical supplies, but also for applications such as aerial surveillance, emergency response, disaster relief missions, passenger transportation, environmental monitoring, and infrastructure inspection in urban areas \cite{gunady2022evaluating}. These novel vehicle categories are expected to play an increasingly significant role in shaping the future of the aviation industry \cite{goyal2021advanced}. With the continuous rise in population density, road transportation is becoming progressively less efficient due to increasing traffic congestion \cite{ikeuchi2019development}. Alternatively, eVTOL emerges as a promising solution, with the ability to operate above congested areas and thereby facilitate smoother and faster deliveries \cite{kim2025low}. These vehicles are still under development and testing phase. However, several challenges remain in AAM development, including safety, infrastructure readiness, airspace integration, cost, and public acceptance. This study focuses on noise because low-altitude eVTOL operations near populated areas may strongly affect community acceptance\cite{rizzi2020urban}. Unlike conventional aircraft, eVTOLs are expected to fly at lower altitudes and over densely populated areas, making it essential to conduct research on the noise impact on people \cite{eissfeldt2022public}. According to a recent survey, the successful integration of eVTOL vehicles depends on effective noise management, a key determinant of public acceptance and closely linked to accurate prediction and mitigation strategies \cite{raza2025noise}. Moreover, eVTOLs will operate from vertiports, which are takeoff and landing hubs for passenger operations. Since vertiports may be located near densely populated areas, noise evaluation is important because frequent operations could increase community noise exposure and affect public acceptance of AAM \cite{yunus2023efficient}. With eVTOL systems anticipated to operate alongside road transportation, accurate noise assessment becomes vital, as these vehicles will generate additional traffic noise in urban areas, necessitating careful consideration by policymakers and stakeholders \cite{schmahl2022semi}. Several studies have examined noise prediction of AAM vehicles. A recent NASA psychoacoustic study found that AAM vehicles and helicopter noise produced similar annoyance responses at equal A-weighted sound exposure levels, although UAM vehicles were generally quieter at equivalent observer distances \cite{vaughn2026helicopter}. In addition, recent review work on AAM broadband noise has shown that different computational, experimental, and data-driven methods are being developed to predict and reduce noise from future AAM vehicles, although further refinement is still needed for reliable community noise assessment \cite{app14188455}. Ground-effect studies further show that UAM noise can increase during vertiport approach, with the maximum overall sound pressure level increasing by more than 3 dBA as the aircraft operates closer to the ground \cite{10.1063/5.0221902}.

However, most research focuses on acoustic metrics (dB/DNL) without extending the analysis to broader societal impacts. While acoustic indicators such as “60 dB” or “65 dB” are standard in engineering analysis, their practical meaning is often not intuitive for communities, making it difficult for policymakers and stakeholders to assess the societal impact of eVTOL operations. These acoustic indicators are perceived differently by each individual, making it challenging to represent the true impact of noise on the community. Prior studies have examined AAM and eVTOL noise mainly from an acoustic perspective. For example, one study developed a grid-based route-planning framework to minimize total and peak sound exposure during UAM operations \cite{doi:10.2514/6.2024-2881}, while another investigated eVTOL-based package delivery planning with community noise impact considerations, including the effects of population exposure, noise-sensitive areas, and time-of-day operations \cite{FARAZI2024103661}. Although these studies advanced noise-aware flight planning, they remained focused primarily on acoustic exposure rather than broader population-based community impact. This paper addresses this gap by developing a framework that translates noise data into human-centered indicators. We apply this framework to the Northwest Arkansas region to evaluate the percentage of the population that is highly annoyed by noise (\%HA). The \%HA metric represents the expected annoyance level of a population, where 0\% indicates no annoyance and 100\% indicates maximum annoyance. This metric can later be used to guide path-planning applications for AAM operations. To implement this framework, we designed a hypothetical route in ArcGIS software. In this paper, we answer the following research questions:
\begin{enumerate} [label=\arabic*.]
    \item 	How can acoustic data from eVTOL operations be converted into \%HA?

\item 	How can spatial estimates of highly annoyed populations guide the optimization of flight routes in AAM systems?

\end{enumerate}

Here, we present a framework that translates eVTOL noise exposure into human annoyance metrics to address the first research question. Expressing results using \%HA provides a clearer way to understand community impact and helps policymakers and engineers evaluate design trade-offs related to public acceptance. This  paper provides further analysis of spatial variation, offering a comprehensive view of how community response can be integrated into the evaluation of AAM operations.

\section{  Study Area and Data Sources }
\vspace{0.1cm}
\noindent\textbf{A. Study Area}
\vspace{0.1cm}

The study is conducted in the Northwest Arkansas region, USA, which represents a rapidly growing urban-suburban environment suitable for evaluating Advanced Air Mobility (AAM) operations. This region is selected due to its increasing population density and expanding healthcare infrastructure, making it a relevant case for eVTOL-based medical delivery applications \cite{nwa_health_vision_2030}.
Since eVTOL networks are still under planning in the United States, existing airports and heliports can serve as initial reference points for mapping \cite{faa2023implementation}. A hypothetical eVTOL route is proposed in this paper. AAM implementation planning is currently focused largely on medical logistics and delivery operations. Therefore, we considered a total of three heliports for the route. Point-A (Washington Regional Medical Center Heliport, Fayetteville, Heliport ID: 4AR8), Point-B (Northwest Medical Center Heliport, Springdale, Heliport ID: 0AR0), and Point-C (Medi-Port Heliport, Rogers, Heliport ID: 27AR). The route was created in ArcGIS Pro, a geographic information system (GIS) software used for spatial mapping and analysis \cite{arcgispro}. A study corridor was defined by extending 3 miles on each side of the route (6 miles total width), representing the region of potential community noise exposure. The distance of the selected route is 15.64 miles. The study area is subdivided into a 1,650-foot grid, and centroids are generated for each cell to measure the spatial distribution of noise and represent population receptor points.

\vspace{0.2cm}
\noindent\textbf{B. Data Sources}
\vspace{0.1cm}

In this study, four main types of data are used: (1) Acoustic noise data from NASA; (2) Human response data from the FAA survey; (3) Population data from the U.S. Census; (4) Spatial mapping data obtained from the Arkansas GIS Office.

The acoustic noise data are obtained from NASA’s third-generation eVTOL Noise–Power–Distance (NPD) database. This data set provides Noise Exposure Level ($L_{AE}$) values for different phases of flight, including climb, cruise, and descent. The NPD data used in this study correspond to the Aviation Environmental Design Tool (AEDT) Helicopter Mode, and results may vary for other eVTOL designs with different acoustic signatures. The term “third-generation” refers to an updated and more advanced set of noise models developed by NASA that incorporate improved aircraft design assumptions and more realistic operational conditions compared to earlier models. These data allow estimation of how noise levels change with distance from the flight path and under different flight conditions \cite{rizzi2023modeling}.

To translate noise exposure into human impact, we used data from the Federal Aviation Administration (FAA) Neighborhood Environmental Survey (NES) \cite{tang2021federal}. This survey provides an exposure-response relationship between the Day-Night Average Sound Level (DNL) and the percentage of highly annoyed individuals
(\%HA), enabling the conversion of physical noise levels into human-centered metrics.

Population data are obtained from the U.S. Census at the block group level to represent how people are distributed across the study area. These data are also used to estimate how many individuals are affected by noise exposure.
Spatial data, including maps of the study area and population regions, are obtained from the Arkansas GIS Office in shapefile format. A shapefile is a standard geospatial data format that stores both location and attribute information for mapping and analysis.
All spatial processing, including grid creation, distance calculation, and overlay analysis, is performed using ArcGIS Pro.
\vspace{-0.45cm}
\section{  Modeling Framework and Spatial Analysis}
\vspace{0.2cm}
\noindent\textbf{A. Overall Workflow}
\vspace{0.1cm}

This study develops a GIS-based framework to estimate human-centered noise impacts from eVTOL operations across the Northwest Arkansas study region. The methodology consists of four integrated stages: (1) acoustic exposure estimation along the route; (2) conversion of acoustic exposure into \%HA; (3) spatial estimation of the number of highly annoyed individuals within each analysis grid; (4) HA-based route evaluation using additional distance- and airspace-related factors to support the identification of optimal flight routes.

The overall workflow begins with route-based noise modeling using NASA acoustic data, followed by conversion of noise exposure into human annoyance using an FAA-based exposure–response function. Census population data are then spatially redistributed into analysis grids, allowing estimation of the highly annoyed population at the grid level. Finally, these HA estimates are subsequently combined with route distance and airspace-related information to evaluate candidate corridors and identify routes that minimize community annoyance impacts.

\vspace{0.2cm}
\noindent\textbf{B. Route and Grid Analysis}
\vspace{0.1cm}

A representative eVTOL route was defined in ArcGIS Pro to connect the selected heliports within the Northwest Arkansas study region. To represent the area potentially affected by aircraft noise, a corridor buffer of 3 miles was created on both sides of the route. This buffered corridor was considered as the study area for subsequent spatial analysis. The route and corresponding study area are shown in Fig. 1.
\begin{figure} [htbp]
    \centering   \includegraphics[width=0.33\linewidth]{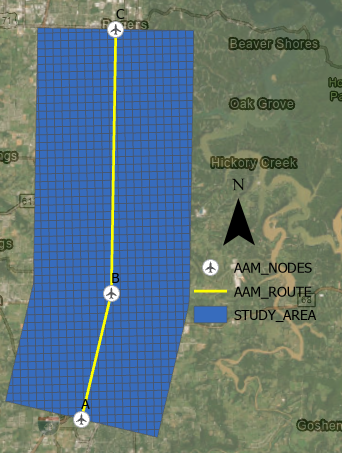}
    \captionsetup{justification=centering}
    \caption{ Northwest Arkansas Study area and eVTOL flight route}
    \label{fig:study_area_route}
\end{figure}
\FloatBarrier
To support spatial analysis, the study area was subdivided into grids. A grid is a rectangular cells that divides the study area into consistent analysis units. In this study, each grid cell was rectangular, allowing noise exposure and population effects to be evaluated systematically across the corridor.
A centroid was assigned to each rectangular grid cell to represent its reference location. These centroids were then used as receptor points for estimating route-related noise exposure. The shortest horizontal distance from each centroid to the eVTOL route was calculated and stored in the attribute table. These distance values were later used to assign noise levels and evaluate how exposure changes with increasing distance from the route.

\vspace{0.2cm}
\noindent\textbf{C. Noise Exposure Estimation}
\vspace{0.1cm}

To characterize route-based acoustic exposure, NASA’s third-generation eVTOL acoustic dataset was used as the primary source of aircraft noise information \cite{rizzi2023modeling}. The dataset provides discrete pairs of horizontal distance and Noise Exposure Level, $L_{AE}$, for different phases of flight, including climb, cruise, and descent. In this study, $L_{AE}$ (dBA) is treated as the single-event aircraft noise metric used to represent the acoustic exposure experienced at a receptor location.

Each rectangular grid cell within the study corridor was represented by its centroid. The shortest horizontal distance from each centroid to the flight route was calculated in ArcGIS Pro. These distances were then used as inputs to the NASA distance-$L_{AE}$ relationship.

Since the centroid distances in the study area did not always exactly match the discrete distance values provided in the NASA acoustic dataset, linear interpolation was applied to estimate $L_{AE}$ at intermediate distances. Specifically, if a centroid distance fell between two known NASA distance points, the corresponding $L_{AE}$ value was estimated by assuming a linear variation between the two neighboring acoustic values. This approach enabled continuous route-based noise estimation across the full study area while preserving the original structure of the NASA acoustic curve.

This interpolation procedure was performed separately for climb, cruise, and descent, yielding phase-specific $L_{AE}$ estimates for all grid centroids. As a result, each centroid was assigned a corresponding noise value, which was subsequently used to compute DNL and human-centered annoyance metrics.

\vspace{0.2cm}
\noindent\textbf{D. Conversion of noise exposure level ($L_{AE}$) to Day-Night Average Sound Level (DNL) and \%HA}
\vspace{0.1cm}

The phase-specific noise exposure values, $L_{AE}$, estimated at each grid centroid were first converted into Day-Night Average Sound Level (DNL), which is a standard cumulative aircraft noise metric used to represent total daily exposure over a 24-hour period. Unlike $L_{AE}$, which describes the sound exposure from a single aircraft event, DNL combines the contribution of all daytime and nighttime operations and applies an additional 10 dB penalty to nighttime events to reflect increased human sensitivity during nighttime hours.

The DNL values were calculated using
\cite{kim2010conversion}. 
\begin{equation}
\mathrm{DNL} = 10 \log_{10} \left[ \sum_{i=1}^{N_d} 10^{0.1 L_{AE}(i)} + \sum_{j=1}^{N_n} 10^{0.1 (L_{AE}(j) + 10)} \right] - 49.4
\end{equation}

\noindent
where, \\
$L_{AE}(i)$: Noise Exposure Level for the $i$-th daytime aircraft event \\
$L_{AE}(j)$: Noise Exposure Level for the $j$-th nighttime aircraft event \\
$N_d$: Number of daytime flights (07:00--22:00) \\
$N_n$: Number of nighttime flights (22:00--07:00, +10 dB penalty applied for each nighttime flight)

\vspace{0.2cm}
Using this relationship, a DNL value was assigned to each centroid in the analysis grid. These centroid-based values were then symbolized spatially across the study area for visualization. Figure 2 shows the resulting spatial distribution of noise expressed as DNL. In this figure, color is used to indicate the magnitude of noise exposure, with darker red tones representing higher DNL values and lighter green tones representing lower values. The figure shows that noise levels are highest near the flight route and gradually decrease as the distance from the route increases. This pattern is expected because aircraft noise weakens as the receiver location becomes farther from the source.

\begin{figure} [htbp]
    \centering   \includegraphics[width=0.35\linewidth]{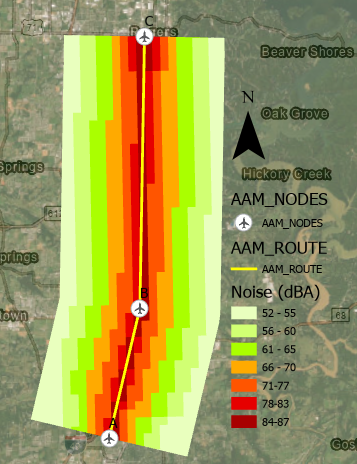}
    \captionsetup{justification=centering}
    \caption{Noise distribution along the eVTOL study area}
    \label{fig:dnl_map}
\end{figure}
\FloatBarrier
After computing DNL, these values were converted into the percentage of highly annoyed individuals (\%HA) using the exposure–response relationship derived from the FAA Neighborhood Environmental Survey (NES) \cite{tang2021federal}. The regression form used in this study is

\begin{equation}
\mathrm{\%HA} = \frac{100\, e^{(b_0 + b_1 * \mathrm{DNL})}}{1 + e^{(b_0 + b_1 * \mathrm{DNL})}} 
\end{equation}
\vspace{-1.9em}
\begin{center}
where, \quad $b_0 = -8.5376,$ \quad $b_1 = 0.1424$
\end{center}

 In this expression, $b_0$ represents the intercept of the fitted relationship and $b_1$ represents the slope, indicating how rapidly annoyance increases with higher DNL values. This function was used to estimate the expected proportion of highly annoyed individuals at each centroid based on its DNL value.

The resulting \%HA values were then mapped across the study area. Figure 3 presents this spatial distribution.

\begin{figure} [htbp]
    \centering   \includegraphics[width=0.350\linewidth]{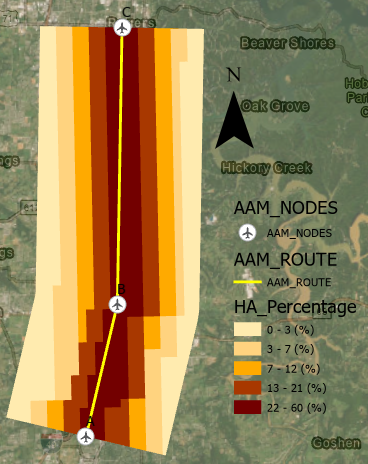}
    \captionsetup{justification=centering}
    \caption{Percentage of highly annoyed (\%HA) heatmap across the study area}
    \label{fig:ha_map}
\end{figure}
\FloatBarrier
\vspace{0.1cm}
Here, the map uses progressively darker shades to indicate higher expected annoyance and lighter shades to indicate lower annoyance. In other words, the figure visually shows where the percentage of highly annoyed individuals is relatively high or low across the corridor. Similar to the DNL pattern, the highest \%HA values occur close to the route, while lower values are observed farther away. This spatial representation provides a more human-centered interpretation of aircraft noise exposure and serves as the basis for the subsequent estimation of the number of highly annoyed individuals in each grid cell.

\vspace{0.2cm}
\noindent\textbf{E. Estimation of Highly Annoyed Population per Grid Cell}
\vspace{0.1cm}

While the percentage of highly annoyed individuals (\%HA) provides a useful human-centered measure of aircraft noise impact, it does not indicate how many people are actually affected in each location. To estimate the number of highly annoyed individuals in each grid cell, population data for Arkansas were obtained from the U.S. Census. In the population shapefile, the study region is divided into irregular census block groups, each with its own total population. However, the noise analysis in this study was performed on a separate set of rectangular grid cells. Because the census block groups and the rectangular analysis grid do not have the same boundaries, the population could not be assigned directly from one dataset to the other.

To address this issue, the census block groups were spatially intersected with the rectangular analysis grid in ArcGIS Pro. This operation divided the original block groups into smaller polygon segments based on the grid boundaries, making it possible to determine how much of each block group lies inside each grid cell. Figure 4 shows the intersected output used in this step.
\begin{figure}[htbp]
    \centering
    \includegraphics[width=0.38\textwidth]{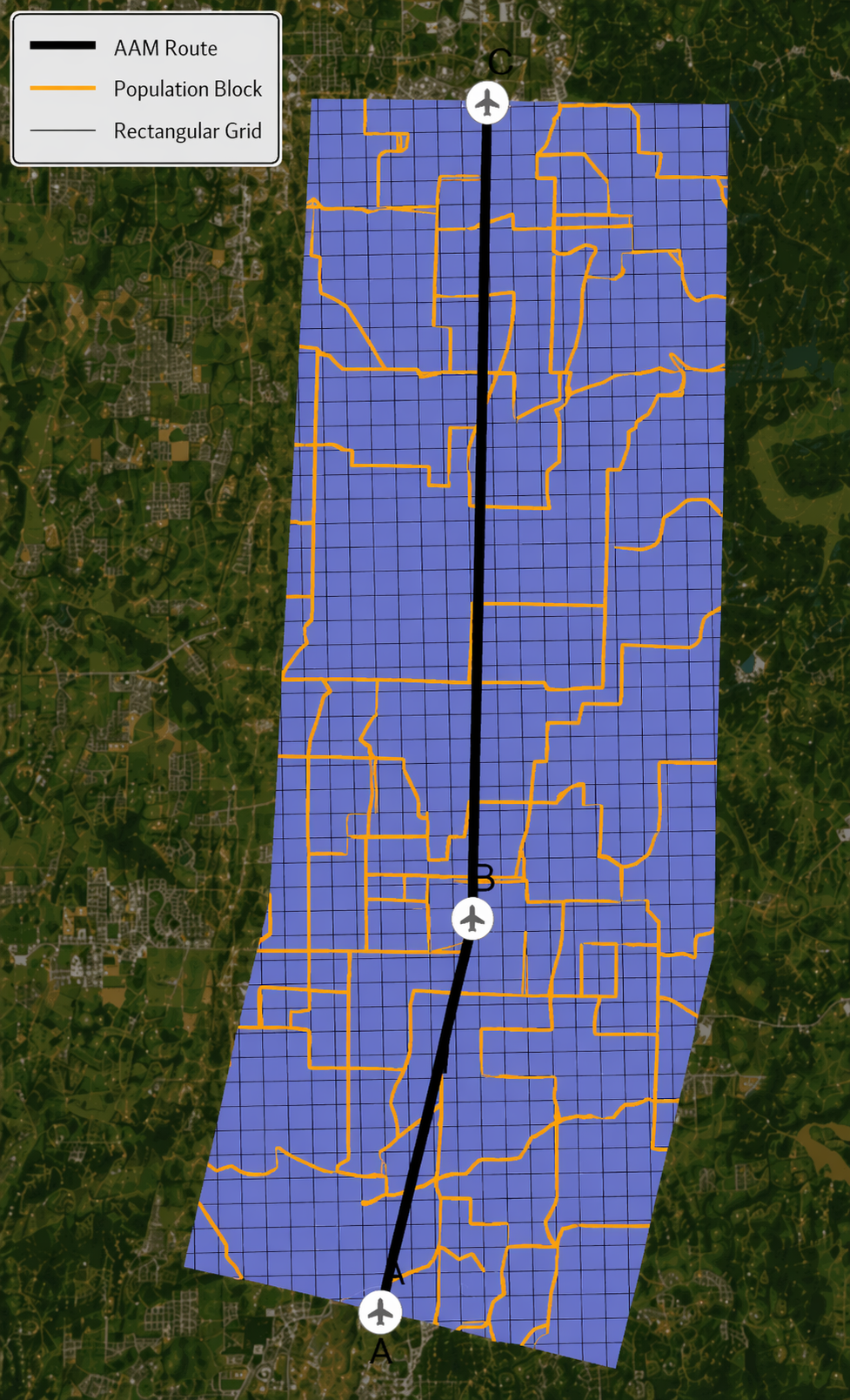}
    \caption{Intersected census block groups and rectangular analysis grid used for area-based population allocation within the study corridor.}
    \label{fig:intersect_grid}
\end{figure}
\FloatBarrier
The population within each grid cell was then estimated to use an area-based allocation approach. Specifically, the portion of a block group’s population assigned to a grid cell was assumed to be proportional to the fraction of the block group area contained within that cell. The grid-level population was calculated as

\begin{equation}
Pop_{grid}=\frac{A_{int}}{A_{BG}}\times Pop_{BG}
\end{equation}

where $A_{int}$ is the area of the block-group portion located inside a grid cell, $A_{BG}$ is the total area of the original census block group, and $Pop_{BG}$ is the total population of that block group.

Because a single grid cell may contain portions of more than one census block group, the population contributions from all intersecting block-group segments were summed to obtain the total estimated population for that grid cell. In this way, the original census population data were redistributed into the rectangular analysis grid so that they could be directly combined with the noise and annoyance estimates.

After estimating the population in each grid cell, the number of highly annoyed individuals was calculated by combining the grid-level population with the corresponding annoyance percentage:

\begin{equation}
HA_{Pop}=Pop_{grid}\times \frac{\%HA}{100}
\end{equation}

where $HA_{Pop}$ represents the estimated number of highly annoyed individuals in a given grid cell.

This step is important because it converts a percentage-based annoyance surface into an estimate of the actual number of affected people. As a result, the analysis moves beyond identifying where annoyance is relatively high and instead identifies where the greatest human impact occurs in terms of population affected.

For visualization, the estimated $HA_{Pop}$ values were displayed as a heatmap over the study corridor, as shown in Fig. 5. In this map, each rectangular grid cell is colored according to the estimated number of highly annoyed individuals within that cell. The legend presents the range of $HA_{Pop}$ values used to classify the grid cells into several groups, from lower values to higher values. The highest number of highly annoyed individuals ranged from 154 to 257 people, while the lowest number ranged from 0 to 8 people. This visualization helps identify the spatial distribution of highly affected areas and provides an intuitive basis for later route assessment and comparison.
\begin{figure}[H]
    \centering    \includegraphics[width=0.45\textwidth]{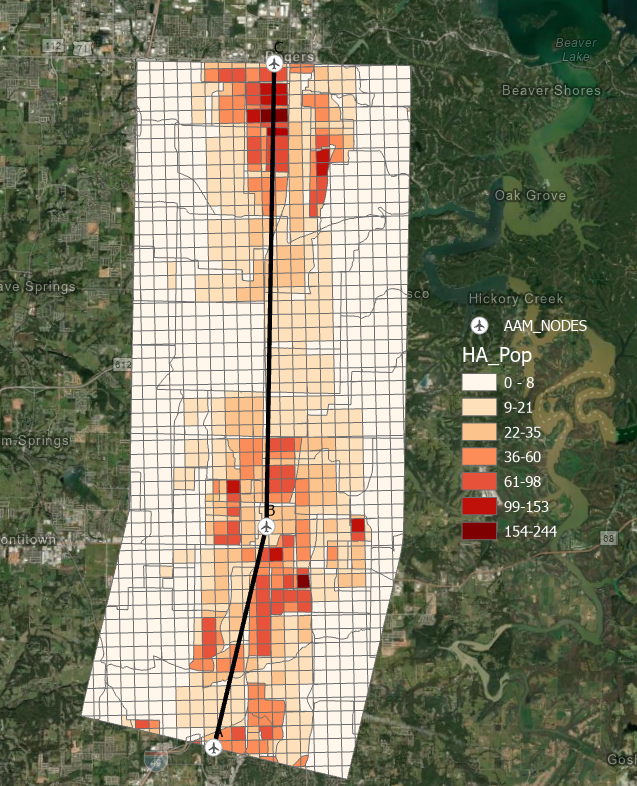}
    \caption{Heatmap of estimated highly annoyed population ($HA_{Pop}$) per grid cell across the study corridor.}
    \label{fig:ha_pop_heatmap}
\end{figure}

\noindent\textbf{F. Multi-Criteria Route Optimization Using Annoyance, Distance, and Airspace Constraints}
\vspace{0.1cm}

After estimating the spatial distribution of the highly annoyed population, the results were further used to optimize alternative eVTOL routes under multiple planning criteria. In this stage, route selection was formulated as a grid-based path optimization problem in which each rectangular grid cell was treated as a node and candidate routes were generated using the A* search algorithm. The A* search algorithm is a pathfinding method that identifies an efficient route between a start point and an end point by comparing the cumulative cost of moving through neighboring grid cells. In this study, route optimization considered three factors simultaneously: travel distance, normalized annoyance, and airspace penalty. For this analysis, the airspace penalty was defined according to the relative operational restrictiveness of FAA airspace classes. Less restrictive Class E and G airspace were assigned a value of 0.1, controlled Class C and D airspace were assigned 0.5, and highly controlled Class B airspace, which typically surrounds the busiest airports, was assigned 1.0. This structure allowed airspace complexity to be included directly in the routing cost function.

Route optimization was performed on a rectangular grid with a cell spacing of 500 m (approximately 0.31 miles), consistent with the 500 m $\times$ 500 m analysis grid used in this study. For each candidate move from node $i$ to node $j$, the edge cost was defined as

\begin{equation}
C_{ij}= \alpha \left(HA_{j}\right)^{p}+
\beta \left(\frac{d_{ij}}{1609.34}\right)+\gamma A_{j}
\end{equation}

where, $HA_{j}$ is the normalized annoyance value associated with the destination cell, $d_{ij}$ is the step distance between neighboring grid cells in meters, $\frac{d_{ij}}{1609.34}$ converts that distance into miles, $A_{j}$ is the airspace penalty value of that cell, and $\alpha$, $\beta$, and $\gamma$ are weighting coefficients representing the relative importance of annoyance, distance, and airspace, respectively. In this study, the exponent p was set to 2.0 to give an extra penalty to cells with high annoyance values, so that the route would more strongly avoid annoyance hotspots. A lower value, such as p=1 or p=0.5, would cause the penalty to increase too slowly and reduce the model’s sensitivity to highly affected cells. On the other hand, a much larger value above p=2 would cause the penalty to increase too sharply, which could lead the route to overreact to a small number of hotspot cells and produce unnecessarily long or impractical paths. Therefore, p=2 was selected as a balanced choice that increases sensitivity to high-annoyance areas without making the routing behavior excessively extreme.

To guide the search toward the destination, a distance-based heuristic term was also included:

\begin{equation}
h(i)=\beta \left(\frac{d_{ig}}{1609.34}\right)
\end{equation}

where $d_{ig}$ is the straight-line distance from the current node to the goal node in meters, and $\frac{d_{ig}}{1609.34}$ converts that distance into miles. This ensured that the search remained distance-aware while still allowing the annoyance and airspace terms to influence the final route.
Seven route configurations were evaluated by varying the weighting coefficients $\alpha$, $\beta$, and $\gamma$, as summarized in Table~\ref{tab:route_weights}.
\begin{table}[H]
\captionsetup{justification=centering, singlelinecheck=false}
\caption{Weighting coefficients used for the seven route configurations.}
\label{tab:route_weights}
\renewcommand{\arraystretch}{1.05} 
\setlength{\tabcolsep}{8pt}
\centering
\begin{tabular}{lccc}
\hline
Route Configuration & $\alpha$ & $\beta$ & $\gamma$ \\
\hline
Distance Only        & 0.0 & 1.0 & 0.0 \\
HA Only              & 1.0 & 0.0 & 0.0 \\
Airspace Only        & 0.0 & 0.0 & 1.0 \\
Distance + Airspace  & 0.0 & 0.7 & 0.3 \\
HA + Airspace        & 0.7 & 0.0 & 0.3 \\
Distance + HA        & 0.7 & 0.3 & 0.0 \\
Combined             & 0.6 & 0.3 & 0.1 \\
\hline
\end{tabular}
\end{table}

The optimized route was generated in two segments, from point A to point B and from point B to point C, and the resulting path was merged into a single route. For each optimized alternative, several route-level metrics were calculated, including total route length, cumulative HA value, maximum HA value, cumulative airspace penalty, and the optimization objective value.

To enable direct comparison among the candidate routes on a common scale, an additional comparison metric was computed as
\begin{equation}
J_{\mathrm{compare}}=W_{HA}HA_{\mathrm{sum}}+W_{\mathrm{DIST}}L_{\mathrm{mi}}+W_{\mathrm{AIR}}A_{\mathrm{sum}}
\end{equation}

where $HA_{\mathrm{sum}}$ is the cumulative normalized annoyance along the route, $L_{\mathrm{mi}}$ is the total route length in miles, and $A_{\mathrm{sum}}$ is the cumulative airspace penalty. In this study, the comparison weights were set to $W_{HA}=0.6$, $W_{\mathrm{DIST}}=0.3$, and $W_{\mathrm{AIR}}=0.1$. Lower values of $J_{\mathrm{compare}}$ indicate more favorable route performance.

This framework made it possible to compare how different routing priorities influence the resulting path and to identify route alternatives that better balance operational efficiency, reduce human annoyance, and airspace compatibility.

\vspace{0.3cm}
\noindent\textbf{G. Sensitivity Analysis for Flight Frequency, Time of Day, and Weather Conditions}
\vspace{0.1cm}

To examine how annoyance outcomes respond to changes in operating conditions, a sensitivity analysis was conducted for flight frequency, time of day, and weather-related factors. Among these, the quantitative scenario analysis in this study focused primarily on changes in the number of daytime and nighttime operations, while time-of-day and weather effects were considered as important influencing factors in the interpretation of annoyance outcomes. Weather effects were not modeled in this study and are identified as an important direction for future work.

For the flight-frequency analysis, one operational scenario were evaluated by changing the number of assumed daytime and nighttime flights in the DNL calculation. The scenario considered 150 daytime flights and 75 nighttime flights. These values were substituted into the DNL formulation given in Eq.(1) to generate scenario-specific cumulative noise exposure at each grid centroid. Time of day was incorporated directly through the DNL framework in Eq.(1), since nighttime operations receive an additional 10 dB penalty compared with daytime operations. Therefore, annoyance outcomes depend not only on the total number of flights, but also on when those flights occur, with nighttime operations contributing more strongly to cumulative annoyance. The updated DNL values were then converted into the percentage of highly annoyed individuals using the same FAA-based exposure–response relationship presented in Eq.(2). 
Finally, the revised annoyance percentages were combined with grid-level population estimates to calculate the corresponding number of highly annoyed individuals under each operating scenario.

Weather was considered an additional sensitivity factor in interpreting annoyance outcomes. A NASA-related study reported that a 15$^\circ$C difference in temperature may have an effect on noise annoyance comparable to a 1--3 dB difference in noise exposure \cite{fields2004meteorological}. In addition, higher wind, humidity, and temperature were associated with a higher likelihood of noise complaints, and they suggested that weather may act as a psychological moderator by altering people’s tolerance and sensitivity to noise \cite{5ec2f1c9-b779-3b7f-b17e-4fc709165fc5}. 

Although weather effects were not explicitly simulated in the present GIS-based framework, these findings indicate that annoyance outcomes may vary not only with route geometry and flight frequency, but also with environmental conditions. Therefore, weather was treated as an important sensitivity factor and a direction for future model refinement. 

\section{  Noise Impact Distribution and Route Assessment}
\vspace{0.2cm}
\noindent\textbf{A. Distance-Based Variation in Percentage of Highly Annoyed Individuals (\%HA)}
\vspace{0.1cm}

The spatial analysis revealed clear spatial variation in acoustic exposure along the eVTOL route. Overall, the heatmaps shown in Fig. 2 and Fig. 3 indicate a strong distance-based reduction pattern consistent with NASA NPD data, highlighting how noise impact and \%HA decrease progressively with increasing distance from the flight route. Fig. 6 shows that the descent phase of flight consistently exhibits higher \%HA values compared to climb and cruise. 

\begin{figure}[H]
    \centering
    \includegraphics[width=0.79\linewidth]{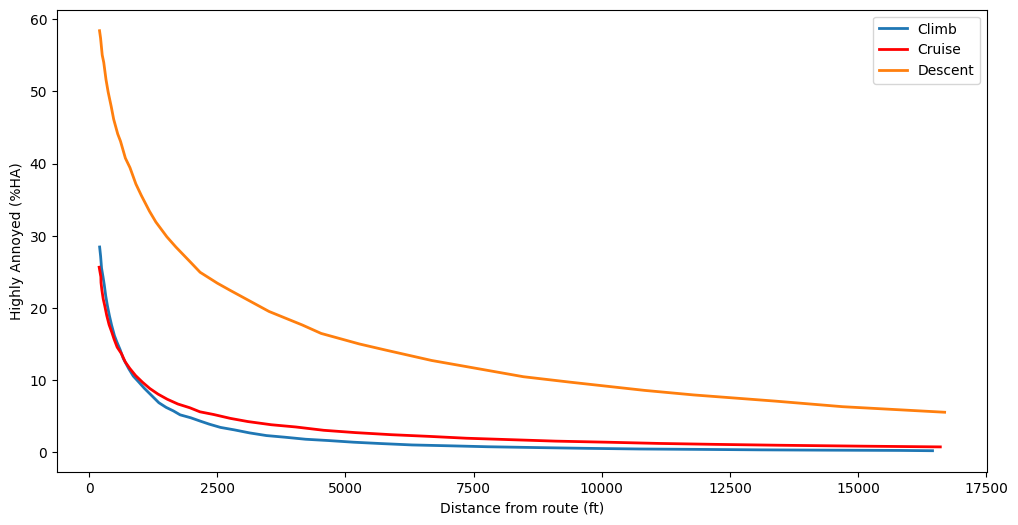}
    \caption{Comparison of \%HA with respect to distance from the route across climb, cruise, and descent phases of flight during eVTOL operations.}
    \label{fig:ha_by_distance}
\end{figure}
Descent phase produced the highest \%HA (30–60\%) within 0 to 1,600 feet distance and decreased to 5.6\% at the end of the study area. During descent, the vehicle operates at lower thrust settings at lower altitudes with rotor noise reaching the ground more directly, resulting in higher exposure levels compared to climb and cruise. The climb phase produced a comparable trend, showing around 29\% HA near the route, which diminished to 0.23\% toward the boundary of the study area. In contrast, the cruise phase demonstrated the lowest annoyance levels, with about 26\% HA near the route, decreasing to 0.55\% at the farthest extent.
Figure 6 illustrates the variation of \%HA with 
distance from the route, which was derived from Day–Night Average Noise Level (DNL) values to provide a more intuitive interpretation for stakeholders. The line chart highlights the expected trend, which is that residents located closer to flight paths experience higher annoyance, while those farther away are less affected.

\vspace{0.5cm}
\noindent\textbf{ B. Identification of the Optimal Route}
\vspace{0.1cm}

Figure 7 compares seven candidate routes over the heatmap of the estimated highly annoyed population. 
\begin{figure}[H]
    \centering
    \hspace*{1.6in}
\includegraphics[width=0.6\textwidth]{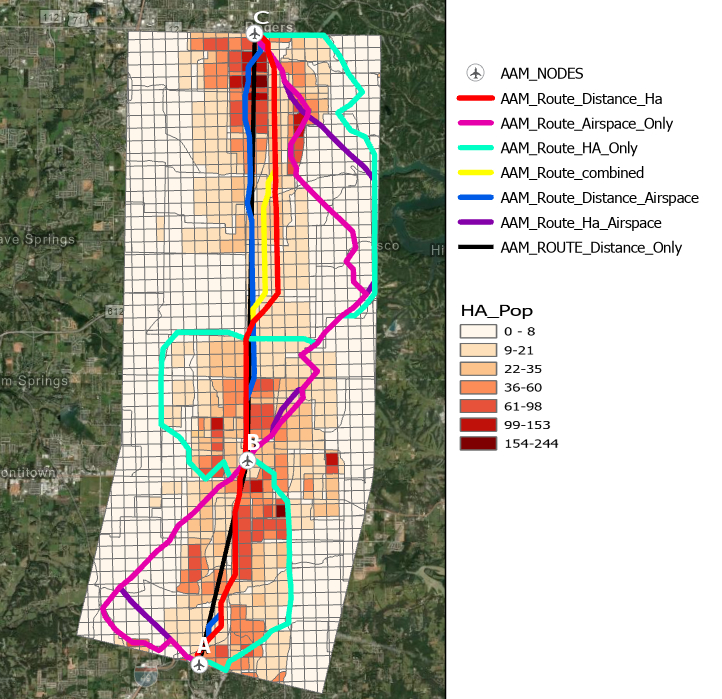}
\captionsetup{justification=centering}
    \caption{Candidate routes overlaid on the estimated highly annoyed population heatmap.}
    
    \label{fig:route_heatmap_op}
\end{figure}

In the background map of Fig. 7, lighter cells indicate lower numbers of highly annoyed people, whereas darker orange and red cells indicate higher affected populations. These higher-impact areas are mainly concentrated near the northern part of the corridor around node C, the central section near node B, and parts of the southern section near node A. The overlaid routes represent the seven optimization cases: Distance Only (black), HA Only (green), Airspace Only (magenta), Distance + HA (red), Distance + Airspace (blue), HA + Airspace (purple), and Combined-[HA+ Distance + Airspace]  (yellow). The figure shows that the candidate routes respond differently to the spatial distribution of the affected population: some remain close to the direct corridor, while others deviate more substantially to avoid the darker hotspot cells.
Table 2 summarizes the route comparison results for the operational scenario (150 daytime and 75 nighttime flights). The candidate routes were compared using total route length, cumulative normalized annoyance, maximum normalized annoyance, airspace penalty, and the comparable total cost, $J_{\mathrm{compare}}$, where lower values indicate better overall performance.

For the 150 daytime and 75 nighttime flights scenario, the route comparison reveals a clear trade-off between operational efficiency and community impact. Routes optimized only for distance remained close to the direct corridor, but they also accumulated the highest annoyance burden, indicating that the shortest path is not necessarily the most desirable from a human-impact perspective. In contrast, the HA Only route achieved the lowest annoyance exposure, but at the cost of a substantially longer path, showing that minimizing annoyance alone can produce routes that are operationally inefficient.

\setcounter{table}{1}
\begin{table}[H]
\renewcommand{\arraystretch}{1.05} 
\setlength{\tabcolsep}{8pt}
\caption{Route comparison results for the 150 daytime and 75 nighttime flights scenario, ordered by increasing total cost ($J_{\mathrm{compare}}$ values).}
\label{tab:route_150_75}
\centering
\resizebox{\columnwidth}{!}{%
\begin{tabular}{lcccccccccc}
\hline
Route & Length (mile) & $HA_{\mathrm{sum}}$ & $HA_{\max}$ & AirPenalty$_{\mathrm{sum}}$ & $J_{\mathrm{compare}}$ & $J_{\mathrm{opt}}$ & $\alpha$ & $\beta$ & $\gamma$ & Cells \\
\hline
Distance + HA        & 16.05 & 3.215 & 0.171 & 24.500 & 12.145 & 7.957  & 0.700 & 0.300 & 0.000 & 51 \\
Combined ( HA+ Distance + Airspace)             & 15.99 & 3.473 & 0.178 & 24.500 & 12.271 & 10.379 & 0.600 & 0.300 & 0.100 & 51 \\
HA + Airspace        & 20.24 & 1.399 & 0.171 & 19.000 & 12.532 & 5.777  & 0.700 & 0.000 & 0.300 & 51 \\
Distance Only        & 15.86 & 5.980 & 0.768 & 24.500 & 13.710 & 25.573 & 0.000 & 1.000 & 0.000 & 51 \\
Distance + Airspace  & 15.86 & 5.980 & 0.768 & 24.500 & 13.710 & 25.251 & 0.000 & 0.700 & 0.300 & 51 \\
Airspace Only        & 21.61 & 2.762 & 0.577 & 19.000 & 14.013 & 19.000 & 0.000 & 0.000 & 1.000 & 54 \\
HA Only              & 27.50 & 0.943 & 0.171 & 33.500 & 17.213 & 0.059  & 1.000 & 0.000 & 0.000 & 77 \\
\hline
\end{tabular}%
}
\end{table}
\vspace{0.03cm}
The Combined route and the Distance + HA route both achieved major reductions in cumulative annoyance relative to the Distance Only route, while requiring only a very small increase in route length. This suggests that a large portion of community exposure can be avoided without substantially compromising travel efficiency. Although the HA + Airspace route further reduced annoyance and airspace penalty, its longer path makes it less practical as a general routing solution.

Overall, Table 2 indicates that the most useful route is not the one that minimizes a single criterion, but the one that best balances competing objectives. From this perspective, the Combined route provides the strongest planning value, because it avoids the most affected population hotspots while maintaining a relatively direct and operationally feasible corridor between nodes A, B, and C.

\section{ Conclusion}
\vspace{0.1cm}

This study developed a GIS-based framework linking acoustic exposure, annoyance modeling, population allocation, and route optimization to evaluate community noise impacts from eVTOL operations across the Northwest Arkansas region. The primary findings addressing the research questions are as follows:

\begin{enumerate}[label=\arabic*., leftmargin=0pt, itemindent=*, labelsep=0.5em]
    \item \textbf{How can acoustic data from eVTOL operations be converted into \%HA?} Acoustic data from eVTOL operations can be converted into \%HA by assigning route-based noise exposure to the analysis grid, converting those values into Day-Night Average Sound Level, and then applying an exposure-response relationship to estimate \%HA. The results showed that \%HA generally decreased with increasing distance from the corridor, while the descent phase produced the highest annoyance levels, reaching approximately 30-60\% near the route. These findings show that acoustic outputs can be translated into a more human-centered indicator that is easier for planners and policymakers to interpret than sound levels alone.

    \item \textbf{How can spatial estimates of highly annoyed populations guide the optimization of flight routes in AAM systems?} Spatial estimates of highly annoyed populations can guide route optimization by identifying where the greatest human impact occurs, rather than only where the highest \%HA values occur. Although some grid cells showed high \%HA, the estimated number of affected people still varied widely across space, ranging from about 0--8 people in some cells to about 154--244 people in others, depending on local population distribution. The Combined and Distance + HA routes both achieved major reductions in cumulative annoyance compared with the Distance Only route, while requiring only a small increase in route length. This shows that route optimization can be guided by avoiding grid cells with larger affected populations, while still accounting for travel distance and airspace constraints. For planners, this provides a practical way to design corridors that reduce community impact in populated hotspots without producing unnecessarily long or operationally unsuitable routes.
\end{enumerate}

Overall, the proposed framework provides a practical, human-centered basis for evaluating eVTOL routes with respect to operational efficiency, airspace compatibility, and community noise impacts. By linking route-level acoustic exposure to population-based annoyance outcomes in a spatially explicit manner, this approach can help planners and policymakers identify noise-sensitive areas, compare alternative corridors, and support more socially acceptable AAM operations. 

Future work should evaluate suitable vertiport locations by identifying sites that minimize highly annoyed population while maintaining operational accessibility. The future research should also incorporate explicit environmental effects, such as temperature and wind, to improve real-world noise and annoyance prediction. The population allocation approach can be improved by accounting for non-uniform population distribution within census block groups, since the present study assumed a uniform distribution, even though the population is more likely to cluster near roads, buildings, and developed areas. Additionally, the framework can be expanded to include other planning and decision-making factors, such as safety, privacy, cost, and automation, to better reflect the broader operational and social challenges of AAM implementation. Further refinement may also include phase-specific operation time, such as the duration of climb, cruise, and descent, to provide a more detailed representation of the eVTOL noise exposure over time.

\section*{Acknowledgments}
\vspace{0.1cm}
This work was supported by the Department of Mechanical Engineering at the University of Arkansas.
\setlength{\bibsep}{0pt}
\vspace{-0.55cm}
\begin{singlespace}
\bibliography{references}
\vspace{0.1cm}
\end{singlespace}

\end{document}